\documentclass[aps,pra]{revtex4}
\usepackage{graphicx}

\begin{document}

\title{Improved W\'{o}jcik's Eavesdropping Attack on Ping-Pong
Protocol Without Eavesdropping-Induced Channel Loss}

\author
{Zhan-jun Zhang$^{1,2,3}$, Yong Li$^{1}$ and Zhong-xiao Man$^2$   \\
{\normalsize $^1$ Department of Physics, Huazhong Normal
University, Wuhan 430079, China} \\
{\normalsize $^2$ Wuhan Institute of Physics and Mathematics,
Chinese Academy of Sciences, Wuhan 430071, China} \\
{\normalsize $^3$ School of Physics \& Material Science, Anhui
University, Hefei 230039, China}}

\maketitle

\begin{minipage}{400pt}
The eavesdropping scheme proposed by W\'{o}jcik [Phys. Rev. Lett.
{\bf 90},157901(2003)] on the ping-pong protocol [Phys. Rev. Lett.
{\bf 89}, 187902(2002)] is improved by constituting a new set of
attack operations. The improved scheme has a zero
eavesdropping-induced channel loss and produces perfect
anticorrelation. Therefore,  the eavesdropper Eve can safely
attack all the transmitted bits and the eavesdropping information
gain can always exceed the legitimate user's information gain in
the whole domain of the quantum channel transmission efficiency
$\eta$, i.e., [0,100\%]. This means that the ping-pong protocol
can be completely eavesdropped in its original version. But the
improvement of the ping-pong protocol security produced by
W\'{o}jcik
is also suitable for our eavesdropping attack. \\

PACS number(s): 03.67.Hk, 03.65.Ud \\
\end{minipage} \\

The private key bits can be produced between two remote parties by
use of quantum key distribution (QKD). QKD is a provably secure
protocol since the security of QKD is based on fundamental laws of
quantum physics. Many theoretical research works [2-20] have been
focused on QKD since the pioneering work of Bennett and Brassard
published in 1984 [1]. These QKD protocols are non-deterministic
since the sender cannot determine the bit value that receiver will
finally decode. Different from this kind of non-deterministic
QKDs, the deterministic secure direct communication protocol is to
transmit directly the secret messages without first generating QKD
to encrypt them. This makes it very useful and usually desired,
especially in some urgent time. However, the deterministic secure
direct communication is more demanding on the security than
non-deterministic QKDs. Therefore, only recently a few of
deterministic secure direct protocols are proposed [21-24]. One of
them is the famous Bostr\"{o}m-Felbinger protocol, or ping-pong
protocol [22], which allows the generation of a deterministic key
or even direct secret communication.

In ping-pong protocol[22,25], Bob prepares two photons in the
entangled state $|\Psi^+ \rangle = (|0\rangle |1\rangle +
|1\rangle |0\rangle ) / \sqrt{2}$. He stores one photon (home
photon) in his lab and sends Alice the other one (travel photon)
via a quantum channel. After receiving the travel photon Alice
randomly switches between the control mode and the message mode.
In the control mode Alice measures the polarization of the travel
photon first in the z-basis and then announces publicly the
measurement result. After knowing Alice's announcement Bob also
switches to the control mode to measure the home photon in the
same basis as that Alice used. Then he compares both measurement
results. They should be perfectly anticorrelated in the absence of
Eve. Therefore, the appearance of identical results is considered
to be the evidence of eavesdropping, and if it occurs the
transmission is aborted. In the other case, the transmission
continues. In the message mode, Alice performs the $Z_t^j  (j \in
\{0,1\})$ operation on the travel photon to encode $j$ and sends
it back to Bob, where $Z=|0\rangle \langle 0| - |1\rangle \langle
1|$. After receiving the travel photon Bob measures the state of
both photons in the Bell basis to decode the $j=0(1)$
corresponding to the $|\Psi^+\rangle(|\Psi^-\rangle)$ result.

The ping-pong protocol has been claimed to be secure and
experimentally feasible[22]. However, since there is a separation
of the verification procedure and the key generation in the
ping-pong protocol, W\'{o}jcik has presented an undetectable
eavesdropping scheme on the Bostr\"{o}m-Felbinger protocol[25].
His eavesdropping attack produces the eavesdropping-induced
channel loss, therefore, requires a lossy channel to work,
otherwise the eavesdropper can be detected by observing the
eavesdropping-induced channel loss. W\'{o}jcik's eavesdropping
scheme produces the eavesdropping-induced channel loss at the
level of 50\% and the anticorrelation of the state of the home
photon (kept by Bob, the legitimate receiver of secret messages)
with that of the travel photon (sent by Bob to Alice, the sender
of secret messages). If the transmission efficiency $\eta$ of the
quantum channel is not taken into account, the probability of the
eavesdropper (i.e., Eve) being detected is zero due to the
anticorrelation. However, in the case of the considerable quantum
channel losses, it is possible for legitimate users to detect the
eavesdropping by observing the quantum channel losses. That is,
although Eve can attack all the transmitted bits and the
eavesdropping-induced channel loss can be hidden in the channel
losses when $\eta \leq 50\%$, if she attacks all the transmitted
bits when $\eta> 50\%$, then the eavesdropping-induced channel
loss is greater than the channel losses and accordingly the
legitimate users can find Eve in the line by observing the channel
losses.

Since W\'{o}jcik's eavesdropping scheme produces the
eavesdropping-induced channel loss at the level of 50\%, it is
desirable to have an eavesdropping scheme on the ping-pong
protocol without eavesdropping-induced channel loss. To construct
his attacking protocol[25], W\'{o}jcik lets Eve use both the
unitary operation and the auxiliary system. since Eve is limited
only by the laws of quantum mechanics, but not at all by current
technology[2], she is free to use any unitary operation. In this
paper, we will improve the W\'{o}jcik's eavesdropping scheme [25]
by using the same auxiliary system but different set of attack
operations. Our improved eavesdropping scheme indeed produces zero
eavesdropping-induced channel loss and never produces both
identical results of the measurements performed by Bob and Alice
in the control mode.

Obviously, Eve has no access to the home photon but can manipulate
the travel photon while it goes from Bob to Alice and back from
Alice to Bob. Eve uses two auxiliary spatial modes $x,y$. She
prepares a photon in the state $|0\rangle$ and lets the other one
be an empty mode, e.g., in the state $|{\rm vac} \rangle_x
|0\rangle_y$. Accordingly, the state of the whole system including
the entangled photon pair is
\begin{eqnarray}
|initial \rangle = |\Psi^+ \rangle_{ht}|{\rm vac} \rangle_x
|0\rangle_y.
\end{eqnarray}
which is the same as W\'{o}jcik's. When Bob sends the travel
photon to Alice, Eve attacks the quantum channel by manipulating
the travel photon through a unitary operation (referred as to be
the $B-A$ attack hereafter) as follow,
\begin{eqnarray}
W_{txy}=U_{txy} V_{txy}Q_{txy},
\end{eqnarray}
where
\begin{eqnarray}
U_{txy} =|0 \rangle \langle 0|_{y}\otimes {\rm
SWAP}_{tx}+(\mathbf{{\cal I} }_{y}-|0\rangle \langle
0|_{y})\otimes \mathbf{{\cal I} }_{tx},
\end{eqnarray}
\begin{eqnarray}
V_{txy} =|1 \rangle \langle 1|_{y}\otimes {\rm
SWAP}_{tx}+(\mathbf{{\cal I} }_{y}-|1\rangle \langle
1|_{y})\otimes \mathbf{{\cal I} }_{tx},
\end{eqnarray}
$\mathbf{{\cal I} }$ is an identity operator,  and $Q_{txy}$ is
defined in [25] as
\begin{eqnarray}
Q_{txy}= {\rm SWAP}_{tx} {\rm CPBS}_{txy} H_y,
\end{eqnarray}
which is composed of the Hadamard gate, the SWAP gate, and the
three-mode gate which is called the controlled polarizing beam
splitter (CPBS) by W\'{o}jcik. When acting on the initial state,
the $B-A$ attack transforms the whole system to the state
$|B-A\rangle= W_{txy}|initial \rangle$ of the form
\begin{eqnarray}
|B-A\rangle= \frac{1}{2}|0\rangle _{h}|1\rangle _{t}(|vac\rangle
_{x}|0\rangle _{y}+|1\rangle _{x}|vac\rangle
_{y})+\frac{1}{2}|1\rangle _{h}|0\rangle _{t}(|vac\rangle
_{x}|1\rangle _{y}+|0\rangle _{x}|vac\rangle _{y}).
\end{eqnarray}

Suppose that Alice now switches to the control mode and measures
the state of the mode $t$.  According to equation (6), we can see
that after the B-A attack Alice will detect with 100\% certainty a
photon whose state is perfectly anticorrelated with the state of
the home photon. Therefore, the probability of eavesdropping
detection based on the correlation observation equals zero. This
point is completely same as that in Ref.[25]. Moreover, from
equation (6) we can also see that the $B-A$ attack in the present
eavesdropping scheme produces zero eavesdropping-induced channel
loss. In contrast, the eavesdropping-induced channel losses in
[25] arrives at 50\%.  This implies that, comparing to the
W\'{o}jcik's eavesdropping scheme, in the present eavesdropping
scheme, the domain in which Eve can attack all the transmitted
bits is enlarged to [0, 100\%] from the [0, 50\%] in Ref.[25].
Nonetheless, this does not certainly mean that the insecurity
upper bound of transmission efficiency presented by W\'{o}jcik can
be pushed up, for now we still do not know the variation of the
eavesdropping (legitimate) information gain. Let us now analyze
the performance of the scheme in the case of Alice operating in
the message mode. After Alice performs the $Z^j$ operation and
sends the travel photon back to Bob, Eve performs her second
attack (named as the $A-B$ attack hereafter) on the travel photon.
The $A-B$ attack consists of the unitary operation $W_{txy}^{-1}$.
After the $A-B$ attack, the corresponding state of the whole
system is
\begin{eqnarray}
|A-B\rangle_j
=\frac{1}{2}[(-1)^j(\Psi^+_{ht}+\Psi^-_{ht})|j\rangle _{y} +
(\Psi^+_{ht}-\Psi^-_{ht})|0\rangle _{y}] |vac\rangle_{x},
\end{eqnarray}
which is little different from the corresponding state in [25] due
to the existence of the partial phase factor $(-1)^j$.

The final step of the eavesdropping scheme is a measurement of
polarization performed on the $y$ photon. The measurement result
is denoted as $k$, while Bob's Bell-state measurement on both
photons is denoted as $m=0 (1)$ corresponding to the
$|\Psi^+\rangle_{ht}(|\Psi^-\rangle_{ht})$ state. Assuming that
Alice sends both values of $j$ with the same probability, then the
only nonzero probabilities $p_{jkm}$ of possible measurement
outputs are
\begin{eqnarray}
p_{000}=1/2, \,\,\, p_{100}=p_{101}=p_{110}=p_{111}=1/8.
\end{eqnarray}
which is the same as W\'{o}jcik's. Therefore, the quantum bit
error rate induced by the eavesdropping in the present
eavesdropping scheme is also at the same level of $1/4$ as that in
Ref.[25]. And the mutual information between any two parties can
be easily worked out,
\begin{eqnarray}
I_{AE}=I_{AB}=\frac{3}{4}\log_2\frac{4}{3} \approx0.311 \nonumber \\
I_{BE}=1-\frac{3}{2}\log_23+\frac{5}{8}\log_25\approx 0.074.
\end{eqnarray}

The same as W\'{o}jcik's eavesdropping scheme, the present scheme
is also not symmetric. If after the $A-B$ attack Eve performs with
the probability of $1/2$ an additional unitary operation $S_{ty}=
X_tZ_t{\rm CNOT}_{ty}X_tZ_t$, where $X_{t}$ is an negation, then
the asymmetry can be removed. If the present $S_{ty}$ is
performed, then the state of the whole system evolves to
\begin{eqnarray}
|A-B\rangle_j^{(S)}= S_{ty} |A-B\rangle_j
=\frac{1}{2}[(\Psi^+_{ht}+\Psi^-_{ht})|j\rangle _{y} +
(-1)^j(\Psi^+_{ht}-\Psi^-_{ht})|1\rangle _{y}] |vac\rangle_{x}.
\end{eqnarray}
Note that the present operation $S_{ty}$ is different from that in
[25]. According to the viewpoint in Ref.[25], that is, since Eve
knows exactly when each of the $S_{ty}$ operations has been
performed, the symmetrization procedure does not reduce the mutual
information between Alice and Eve while it disturbs the
communication between Alice and Eve in such a way the mutual
information between Alice and Bob is reduced. In the present
scheme after the symmetrization procedure the mutual information
between Alice and Bob is also reduced to
$I_{AB}=\frac{3}{4}\log_23-1 \approx0.189$.

Thus far, we have improved W\'{o}jcik's eavesdropping scheme by
constructing a new set of attack operations. We can see that the
improved scheme is almost the same as W\'{o}jcik's eavesdropping
scheme except for the eavesdropping-induced channel loss. Hence,
all the discussions in Ref.[25] except for those related to the
eavesdropping-induced channel loss are also suitable for the
present paper and we will not repeat them. Now let us discuss a
very important property related to the eavesdropping-induced
channel loss. In the improved scheme the $\eta$ domain in which
Eve can attack all the transmitted bits is enlarged to [0, 100\%]
from [0, 50\%] in Ref.[25]. In the $\eta$ domain of (50\%, 100\%],
the eavesdropping information gain does not decrease in the
improved scheme but it does decrease in W\'{o}jcik's eavesdropping
scheme. Thus, in this sense, we can say that the present zero
eavesdropping-induced channel loss does induce more eavesdropping
information gain. Hence, the ping-pong protocol in its original
version is insecure and can be completely eavesdropped, even in an
ideal channel. Although in [22] the security proof against
eavesdropping is provided, we think, it is essentially not a
general proof but only the result of a special choice of the
ancilla. Our present eavesdropping scheme employing two ancillas
from W\'{o}jcik instead of one ancilla in [22] has shown the
insecurity of ping-pong protocol.

In summary, we have improved W\'{o}jcik's eavesdropping scheme on
the Bostr\"{o}m-Felbinger quantum communication protocol. The
improved scheme does not produce any eavesdropping-induced channel
loss and accordingly in the whole $\eta$ domain of [0, 100\%] Eve
can attack all the transmitted bits. Hence, in the $\eta$ domain
of (50\%,100\%], the zero eavesdropping-induced channel loss does
induce more eavesdropping information gain. Moreover, as for as
the original ping-pong protocol is concerned, it is insecure and
can be completely eavesdropped, even in an ideal channel, for the
eavesdropping information gain can always exceed the legitimate
user's information gain. As mentioned by W\'{o}jcik[25], the
ping-pong protocol can be made secure against his attack scheme by
a modification. His improvement of the ping-pong protocol security
is also suitable for our eavesdropping attack. Actually there is a
significant chance of finding an additional photon in the travel
mode in our attack, which can then be used as an indicator for
Eve's presence. The present eavesdropping scheme introduces a
qubit error rate at the same level of 25\% as one for W\'{o}jcik's
attack scheme. If the channel has itself a QBER significantly
lower than 25\%, and if Alice and Bob sacrifice some of their
message bits to perform message authentification, the qubit error
rate induced by the attack can be used to detect the eavesdropper.
Furthermore, another improvement of the ping-pong protocol is
that, if both Alice and Bob use the generally used
two-measuring-basis method [23] in the control mode, our present
eavesdropping attack can also be detected with 25\% possibility.

We thank to Prof. Baiwen Li for his encouragement. This work
is funded by the National Science Foundation of China under No.10304022.\\

\noindent [1] C. H. Bennett and G. Brassard, in {\it Proceedings
of the IEEE International Conference on Computers, Systems and
Signal Processings, Bangalore, India} (IEEE, New York, 1984),
p175.

\noindent[2] N. Gisin, G. Ribordy, W. Tittel, and H. Zbinden, Rev.
Mod. Phys. {\bf 74},145 (2002).

\noindent[3] A. K. Ekert, Phys. Rev. Lett. {\bf67}, 661 (1991).

\noindent[4] C. H. Bennett, Phys. Rev. Lett. {\bf68}, 3121
 (1992).

\noindent[5] C. H. Bennett, G. Brassard, and N.D. Mermin, Phys.
Rev. Lett. {\bf68}, 557(1992).

\noindent[6] L. Goldenberg and L. Vaidman, Phys. Rev. Lett.
{\bf75}, 1239  (1995).

\noindent[7] B. Huttner, N. Imoto, N. Gisin, and T. Mor, Phys.
Rev. A {\bf51}, 1863 (1995).

\noindent[8] M. Koashi and N. Imoto, Phys. Rev. Lett. {\bf79},
2383 (1997).

\noindent[9] W. Y. Hwang, I. G. Koh, and Y. D. Han, Phys. Lett. A
{\bf244}, 489 (1998).

\noindent[10] P. Xue, C. F. Li, and G. C. Guo,  Phys. Rev. A
{\bf65}, 022317 (2002).

\noindent[11] S. J. D. Phoenix, S. M. Barnett, P. D. Townsend, and
K. J. Blow, J. Mod. Opt. {\bf42}, 1155 (1995).

\noindent[12] H. Bechmann-Pasquinucci and N. Gisin, Phys. Rev. A
{\bf59}, 4238 (1999).

\noindent[13] A. Cabello, Phys. Rev. A {\bf61},052312 (2000);
{\bf64}, 024301 (2001).

\noindent[14] A. Cabello, Phys. Rev. Lett. {\bf85}, 5635 (2000).

\noindent[15] G. P. Guo, C. F. Li, B. S. Shi, J. Li, and G. C.
Guo, Phys. Rev. A {\bf64}, 042301 (2001).

\noindent[16] G. L. Long and X. S. Liu, Phys. Rev. A {\bf65},
032302 (2002).

\noindent[17] F. G. Deng and G. L. Long, Phys. Rev. A {\bf68},
042315 (2003).

\noindent[18]J. W. Lee, E. K. Lee, Y. W. Chung, H. W. Lee, and J.
Kim, Phys. Rev. A {\bf 68}, 012324 (2003).

\noindent[19] Daegene Song, Phys. Rev. A {\bf69}, 034301(2004).

\noindent[20] X. B. Wang, Phys. Rev. Lett. {\bf 92}, 077902
(2004).

\noindent[21] A. Beige, B. G. Englert, C. Kurtsiefer, and
H.Weinfurter, Acta Phys. Pol. A {\bf101}, 357 (2002).

\noindent[22] Kim Bostrom and Timo Felbinger, Phys. Rev. Lett.
{\bf89}, 187902 (2002).

\noindent[23] F. G. Deng, G. L. Long, and X. S. Liu, Phys. Rev. A
{\bf68}, 042317 (2003).

\noindent[24]F. G. Deng and G. L. Long,  Phys. Rev. A {\bf 69},
052319 (2004).

\noindent [25] A. Wojcik, Phys. Rev. Lett. {\bf90}, 157901 (2003).

\end{document}